\pdfoutput=1

\documentclass[3p,times,procedia]{elsarticle}
\usepackage{nupha_ecrc}
\usepackage{subcaption}

\usepackage{amsmath}

\volume{00}

\firstpage{1}

\journalname{Nuclear Physics A}

\runauth{}

\jid{nupha}

\jnltitlelogo{Nuclear Physics A}

\usepackage{amssymb}

\usepackage[figuresright]{rotating}
\graphicspath{{images/}}

\begin{document}

\begin{frontmatter}



\dochead{}

\title{Measurement of D-meson azimuthal anisotropy in Au+Au 200 GeV collisions at RHIC}


\author{Michael Lomnitz, for the STAR collaboration}

\address{Lawrence Berkeley National Laboratory, One Cyclotron Road, MS 70R0319, Berkeley CA, 94720}
\address{Kent State University, Physics Department, 800 E. Summit St., Kent, OH 44240}

\begin{abstract}
Heavy quarks are produced through initial hard scatterings and they are affected by the hot and dense medium created in heavy-ion collisions throughout its whole evolution. Due to their heavy mass, charm quarks are expected to thermalize much more slowly than light flavor quarks. The charm quark flow is a unique tool to study the extent of thermalization of the bulk medium dominated by light quarks and gluons. At high $p_T$, D meson azimuthal anisotropy is sensitive to the path length dependence of charm quark energy loss in the medium, which offers new insights into heavy quark energy loss mechanisms - gluon radiation vs. collisional processes.\\
We present the STAR measurement of elliptic flow ($v_2$) of $D^0$ and $D^{\pm}$ mesons in Au+Au collisions at $\sqrt{s_{NN}}$ = 200 GeV, for a wide transverse momentum range. These results are obtained from the data taken in the first year of physics running of the new STAR Heavy Flavor Tracker detector, which greatly improves open heavy flavor hadron measurements by the topological reconstruction of secondary decay vertices. The D meson $v_2$ is finite for $p_T > 2 \text{ GeV/c}^2$ and systematically below the measurement of light particle species at the same energy. Comparison to a series of model calculations favors scenarios where charm flows with the medium and is used to infer a range for the charm diffusion coefficient $2\pi T D_s$.
\end{abstract}


\end{frontmatter}


\section{Introduction}
  Heavy flavor quarks are suggested to be an excellent probe to study the strongly coupled quark-gluon plasma (sQGP) as they are produced early in heavy-ion collisions through hard scattering processes and experience the full evolution of the system, while their large masses are mostly unaffected by the QCD medium.  Furthermore, drawing an analogy to Brownian motion the heavy quarks propagating through the medium are sensitive to the sQGP transport properties, for example $2\pi T D_s$, which is given in term of the temperature $T$ and the charm diffusion coefficient $D_s$ \cite{Abreu20112737}.\\
  Recent measurements at RHIC and LHC show that high $p_T$ charmed hadron yields in central collisions are considerably suppressed suggesting strong interactions, while the elliptic flow $v_2$ measured at LHC is comparable to that of light hadrons \cite{PhysRevLett.113.142301,PhysRevLett.111.102301,arXiv.1509.06888}. At RHIC, the enhancement in the nuclear modification factor at intermediate $p_T$ is suggestive of both charm flow and production via coalescence. However, charm flow inferred from measurements of semi-leptonic decays suffer from large uncertainties. A precise measurement of $v_2$ over a broad $p_T$ range, and in particular at low momenta, will provide useful insights into the properties of sQGP medium.
 
\section{Experimental setup}
  The data used in this analysis were recorded in year 2014 by the STAR experiment at the Relativistic Heavy ion Collider (RHIC) in Brookhaven National Laboratory, USA.  The STAR experiment possesses full azimuthal coverage at mid-rapidity using the Time Projection Chamber (TPC) to reconstruct tracks inside a uniform 0.5 T magnetic field.  The entire Heavy Flavor Tracker (HFT) micro-vertexing detector was included for the first time in 2014 and greatly improved STAR's tracking resolution, providing track pointing resolution of less than $50\mu$m for kaons with $p_T=750$ MeV/c.\\
  About 780 million Au+Au collisions at $\sqrt{s_{NN}}$ = 200 GeV events recorded with a Minimum Bias (MB) trigger in 2014 were analyzed to reconstruct charmed hadrons. A cut on the reconstructed collision position along the beam line ($|$ primary vertex $|< 6$ cm) is applied to ensure good detector acceptance.\\ 
$D$-mesons are reconstructed in the hadronic channels:\\

 \begin{tabular}{ l r }
 $D^0(\overline{D^0})\rightarrow K^\mp \pi^\pm$, $c\tau \sim 120$ $\mu m$ B.R. 3.9\% &
 $D^\pm\rightarrow K^\mp 2\pi^\pm$, $c\tau \sim 300$ $\mu m$ B.R. 9.1\%
 \end{tabular}\\
 
 Daughter tracks are required to have a minimum of 20 hits in the TPC and hits in all three layers of the HFT, $p_T > 0.6$ GeV/c and pseudorapidity $|\eta|<1$.  Particle identification is done using energy loss dE/dx from the TPC, selecting candidates within 2 to 3 standard deviations from the expected value and is further enhanced by use of the Time of Flight detector(TOF) when available. $1/\beta$ is estimated from the momentum and timing from TOF, and is required to be less than 0.03 from the expected value.\\
 Once daughter candidates have been identified, the decay vertex can be reconstructed, which is displaced from the primary vertex of collision. In the case of two body decays the decay is reconstructed at the mid point on their distance of closest approach (DCA). For three body decays, such as $D^\pm$, the average between the midpoints of pairwise DCA's is taken as the decay position.\\
 The combinatorial background can be greatly suppressed by cutting on the following topological variables: decay length (distance between primary and decay vertices), DCA between daughter tracks, DCA between reconstructed parent and the primary vertex (PV), DCA between daughter tracks and the PV; and in the case of $D^\pm$ the distances between the midpoints from each pairwise combination.
  \section{Azimuthal anisotropy}
\begin{figure}
 \centering
 \begin{subfigure}[t]{0.3\textwidth}
 \includegraphics[width=\linewidth]{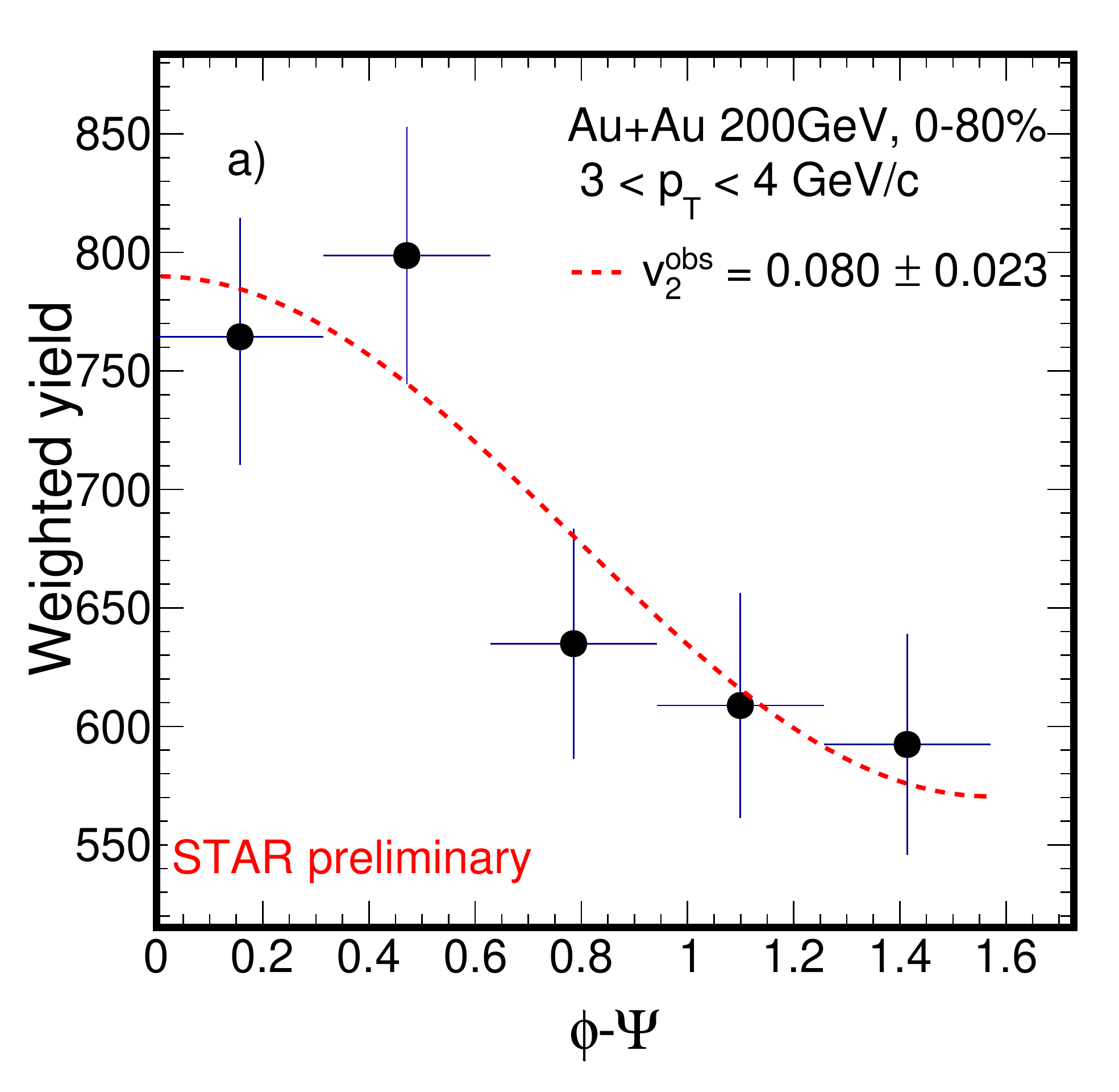}
 \phantomcaption{}\label{phiYield}
 \end{subfigure} 
  \begin{subfigure}[t]{0.3\textwidth}
 \includegraphics[width=\linewidth]{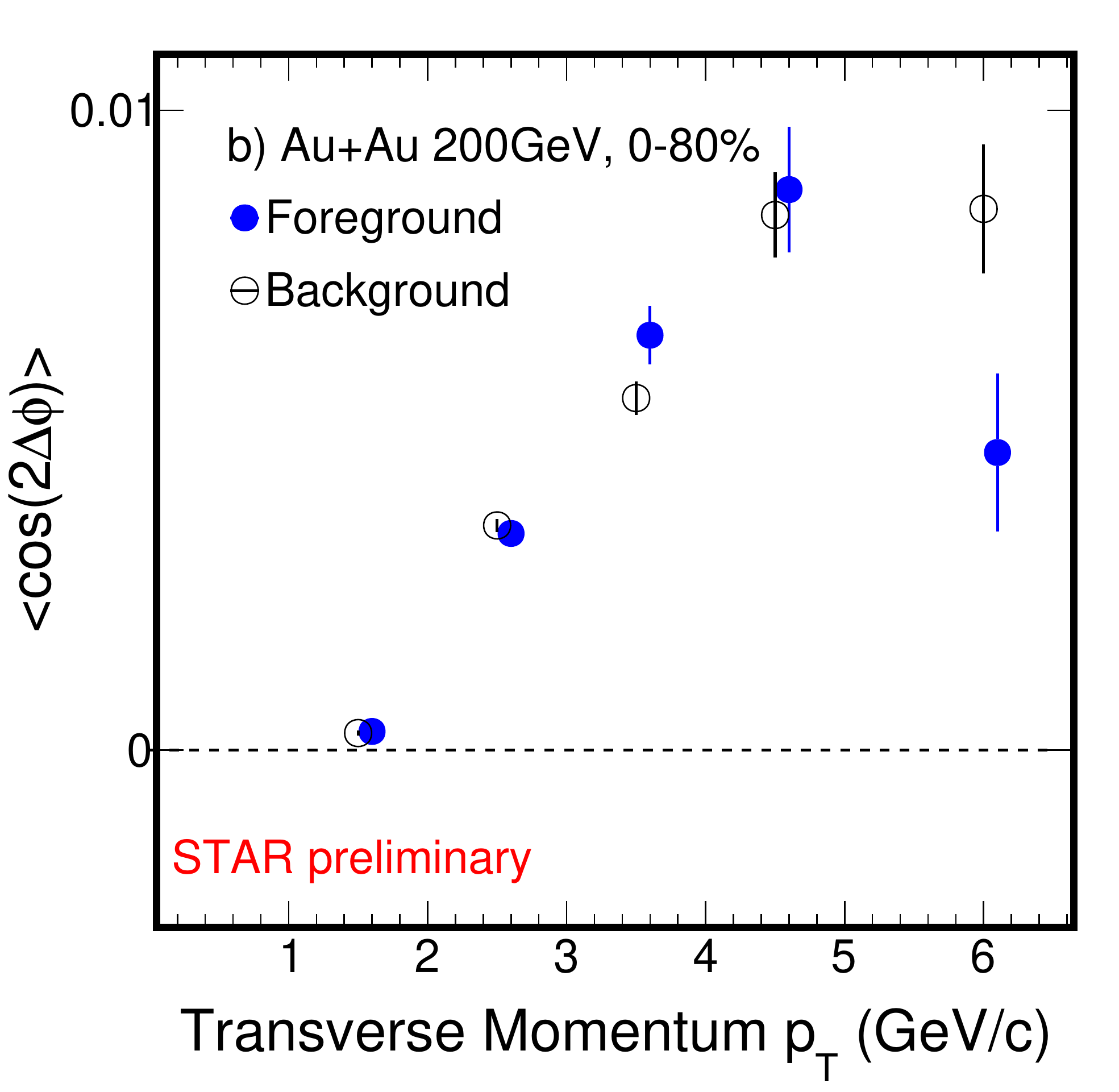}
  \phantomcaption{}\label{2pcV2}
 \end{subfigure} 
  \begin{subfigure}[t]{0.3\textwidth}
 \includegraphics[width=\linewidth]{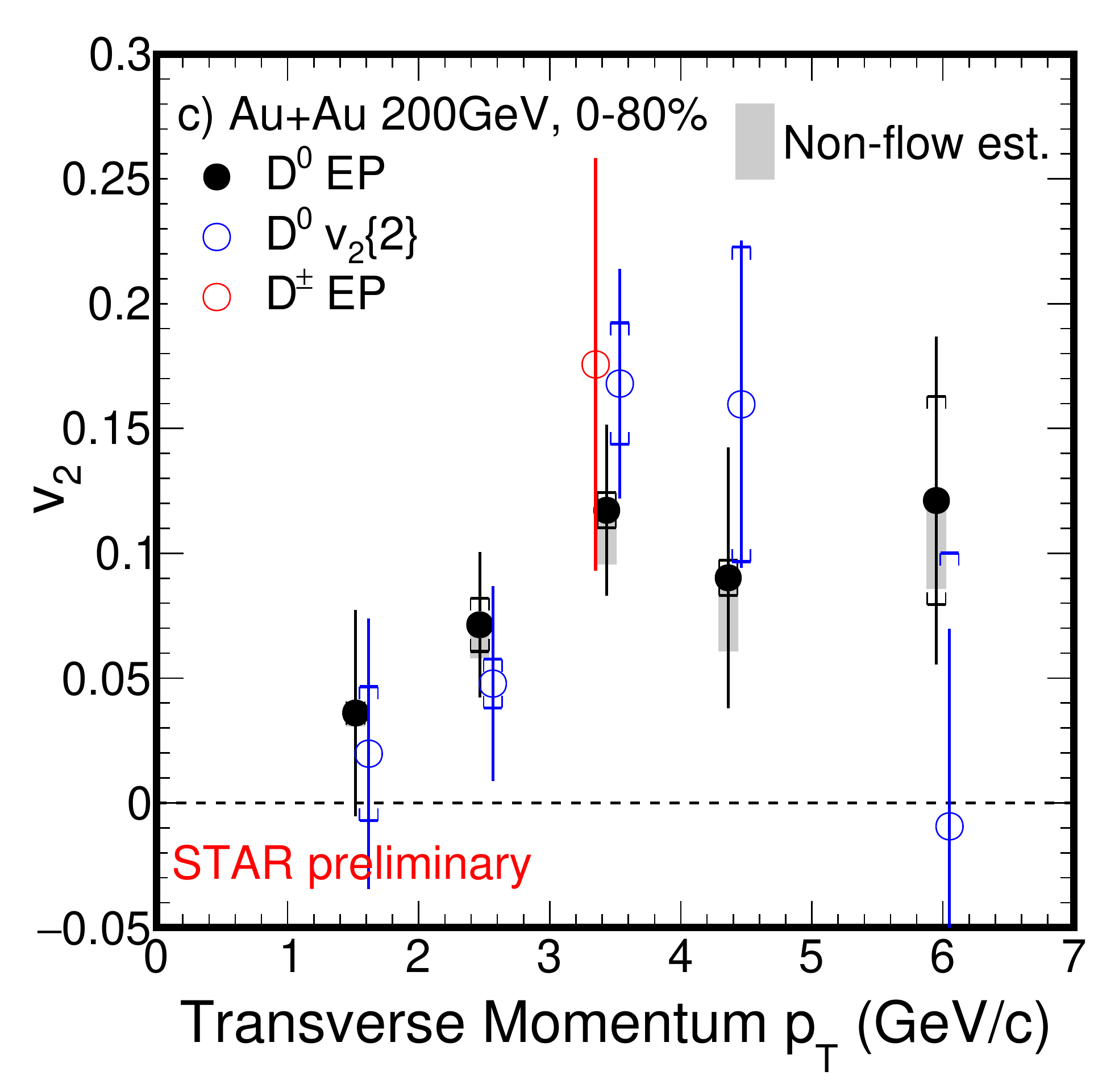}
\phantomcaption{}\label{methodsv2}
 \end{subfigure}
\caption{a) Weighted $D^0$ yield vs. $\phi-\Psi$ for 0-80\% central collisions and $p_T\in[3,4]$ GeV/c. b) $<\cos(2\Delta\phi)>$ between (un)like-sign pairs and charged particles as a function of $p_T$. c) Comparison between $v_2$ from event plane and two particle correlation methods.}
\label{fig:v2Methods}
\end{figure}

\begin{figure}
 \centering
  \begin{subfigure}[t]{0.3\textwidth}
 \includegraphics[width=\linewidth]{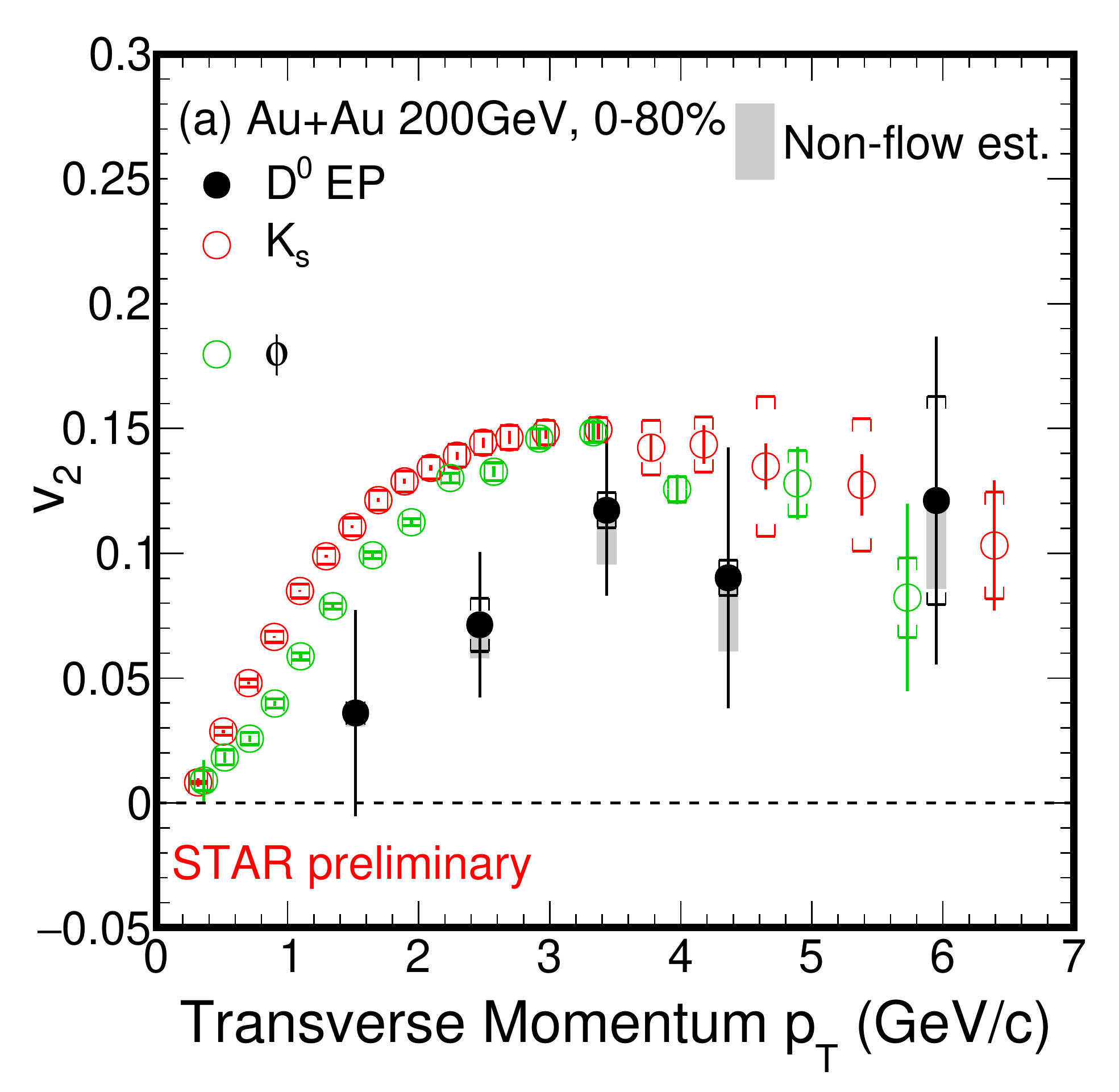}
\phantomcaption{}\label{speciesv2}
 \end{subfigure} 
 \begin{subfigure}[t]{0.3\textwidth}
 \includegraphics[width=\linewidth]{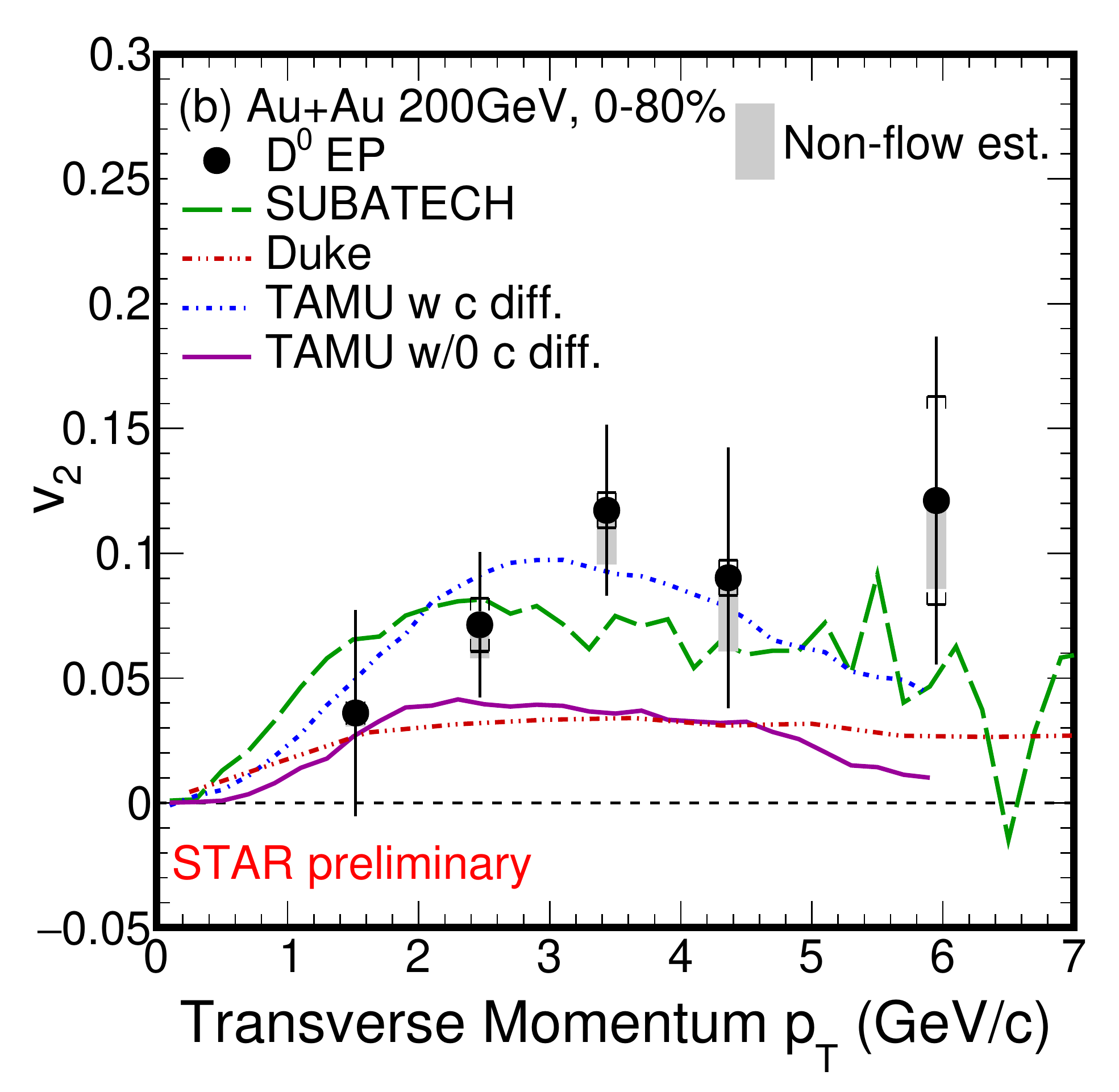}
 \phantomcaption{}\label{theoryv2}
 \end{subfigure} 
  \begin{subfigure}[t]{0.35\textwidth}
 \includegraphics[width=\linewidth]{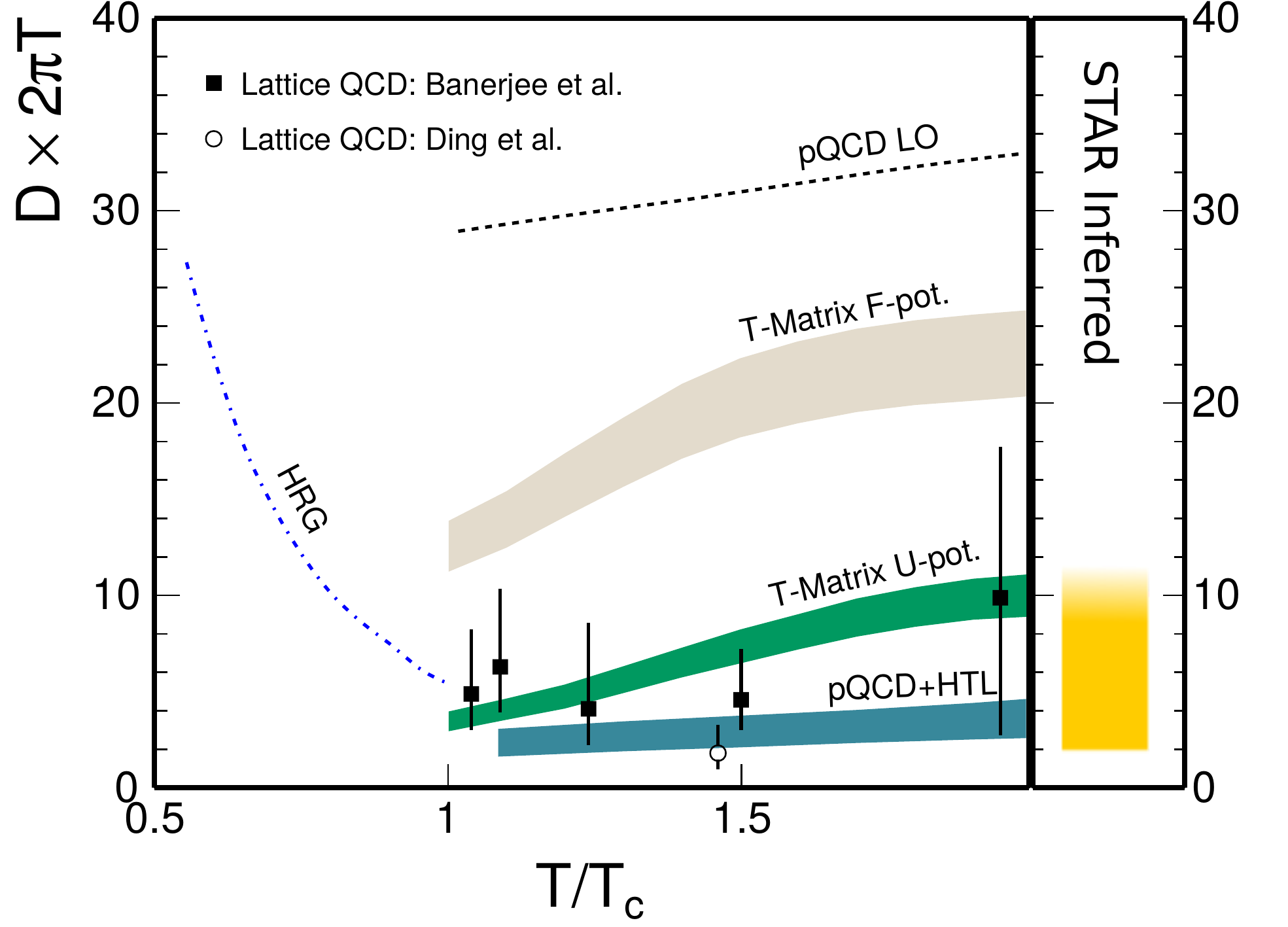}
 \phantomcaption{}\label{diffusionCoef}
 \end{subfigure} 
  \put(-130,110){\footnotesize (c)}
\caption{a) Measured $v_2$ for $D^0$ compared to that of light hadrons. b) Comparison for measured $D^0$ $v_2$ and model calculations. c) Diffusion coefficient from model calculations and inferred range from STAR results.}
\end{figure}

  Once $D$-meson candidates have been selected, the second order azimuthal anisotropy, $v_2$, is studied using two different methods: the event plane method and the two particle correlation method, which will be discussed briefly in the following paragraphs.\\
   In the event plane method the second order event plane, $\Psi$, is reconstructed from TPC tracks and corrected for the non-uniform detector efficiency \cite{v2Methods}. In order to reduce the non-flow contributions from other two or multi-body correlations, a relative pseudorapidity gap $|\Delta\eta| \leq 0.15$ around $D^0$ candidates is excluded from the event plane reconstruction. The azimuthal distribution of $D$ mesons with respect to the event plane $\phi-\Psi$ is then obtained and weighted by $1/\epsilon/R$, the inverse of the $D^0$ reconstruction efficiency $\epsilon$ and the event plane resolution R \cite{arXiv.1212.3650} for each centrality.\\
   In each $\phi - \Psi$ bin the mixed-event background is scaled to the like-sign background and subtracted from the unlike-sign invariant mass spectrum. The $D^0$ yield is obtained by either the fit or sideband method: at low $p_T$ the invariant mass spectrum is fitted with a Gaussian, representing the signal, and a first order polynomial describing the correlated background; for the last $p_T$ bin (5-10 GeV/c) the fit is limited by low background statistics, and the $D^0$ yield is obtained by subtracting scaled counts in two invariant mass regions around the signal region.\\
  The observed $v_2^{obs}$ is then obtained by fitting the yield versus $\phi - \Psi$ with the functional form $A(1+2v_2\cos(2(\phi-\Psi)))$. Finally, the observed $v^{obs}_2$ is corrected for the average even plane resolution $<1/R>$ to obtain the true value of $v_2$. Figure \ref{phiYield} shows the weighted yield as a function of the angle relative to the event plane $\phi - \Psi$ for $D^0$ candidates with $3< p_T < 4$ GeV/c.\\
  In the two particle correlation method, the average $D^0$-hadron correlation is calculated $V_2^{D-h} \equiv <\cos(2\phi_D - 2 \phi_h)>$. Assuming that the $D^0$ and hadron have no correlation other than with the event plane, it can be shown that $V_2^{D-h}  = v_2^Dv_2^h$. Finally, together with the hadron-hadron correlation, $V_2^{h1-h2} = (v_2^h)^2$,  the $D$-meson $v^D_2 = <\cos(2\phi_D - 2 \phi_h)>/(<\cos(2\phi_{h1} - 2 \phi_{h2})>)^{1/2}$ can be obtained. $D^0$ background is estimated by the average of the sidebands (both like-sign and unlike-sign) as well as the $D^0$ like-sign invariant mass spectrum in the signal range. Figure \ref{2pcV2} shows $V_2$ for $D^0$ candidates and background (within the $D^0$ invariant mass range) as a function of transverse momentum.\\
  As in the case of the event plane method, the contribution from non-flow is suppressed by introducing a gap $| \Delta\eta | > 0.2$ in the measurement of  $D-h$ correlations, and particles in opposite sides of the TPC ($h_1\in \eta < 0$ and $h_2\in\eta > 0$) are used for the hadron-hadron correlation.\\
  Due to limited statistics, $D^\pm$ yield is obtained in and out-of plane and the $v_2$ is obtained following the event plane method.  Figure \ref{methodsv2} shows the azimuthal anisotropy parameter for both $D^0$ and $D^\pm$ as well as a comparison between both methods. The event plane and two particle correlation results show good overall agreement within systematic uncertainties, and the $D^\pm$ $v_2$ is consistent, within large uncertainties, with that of $D^0$.\\
  The remaining contribution of non-flow effects to the measured $v_2$ is estimated by scaling the non-flow in p+p collisions to Au+Au \cite{PhysRevLett.93.252301}. 
  In this case the non-flow contribution can be written as $\left< \sum_i \cos(2(\phi_{D^0} - \phi_h))\right> / M\overline{v_2}$, where the sum $i$ is done over near side charged hadrons in p+p, $M$ and $\overline{v_2}$ are the average multiplicity and charged hadron $v_2$ in Au+Au.\\
  Figure \ref{speciesv2} shows the $v_2$ for $D$-mesons compared with other particle species.  The $D^0$ azimuthal anisotropy is significantly different from zero for $p_T>2$ GeV/c and is systematically lower than that of the lighter hadrons \cite{PhysRevC.77.054901} in the range for $1<p_T<4$ GeV/c, but the extent of heavy flavor thermalization is still under investigation.
   In figure \ref{theoryv2} the results for $D^0$ are compared to four theoretical models. The calculation by the SUBATECH group \cite{PhysRevC.91.014904} employs pQCD with Hard Thermal Loop approximation for soft collisions and agrees with the measurement over the whole $p_T$ range. The TAMU model \cite{PhysRevC.86.014903} uses a non-perturbative T-matrix approach assuming the two-body interactions can be described by a potential as a function of the transfered 4-momentum. Two curves from TAMU are shown: the scenario including charm diffusion(blue) agrees with the data while the predictions without charm diffusion(magenta) are consistently lower.  In the model developed by the Duke university group \cite{PhysRevC.88.044907} the diffusion coefficient $2\pi T D_s$ is a free parameter which, in the case of the red curve shown, has been constrained using the $R_{AA}$ measured at LHC with a value of roughly 7, and under-predicts the $v_2$ observed in our data. 
  Figure \ref{diffusionCoef} shows the value obtained for the diffusion coefficient from different model calculations compared to the yellow band on the far right showing the range of inferred values that are compatible with the measurement from STAR.
     
\section{Summary}
 STAR has carried out the first heavy flavor measurements in heavy ion collisions using the newly installed, state of the art vertexing detector, the HFT. The measured charmed meson $v_2$ in Au+Au collisions is found to be finite, though systematically below $v_2$ of light hadrons. Comparison to a series of models shows that they are able to describe the data, favoring the scenario where charm quarks flow with the medium. The models infer a range of compatible values for the charm diffusion coefficient $2\pi T D_s$ between  2 or 10.\\
 As part of STAR's continuing heavy flavor program, 2 billion Au+Au minimum bias events will be recorded in 2016 with full aluminum cables in the innermost silicon layer of the HFT. An expected 2-3 factor improvement in the $D^0$ significance at $p_T = 1$ $\mathrm{GeV/c}^2$ will allow the study of the centrality dependence of $v_2$ in heavy ion collisions.
\section{Acknowledgements}
This material is based upon work supported by the U.S. Department of Energy, Office of Science, Office of Workforce Development for Teachers and Scientists, Office of Science Graduate Student Research (SCGSR) program. The SCGSR program is administered by the Oak Ridge Institute for Science and Education for the DOE under contract number DE-AC05-06OR23100.




\bibliographystyle{elsarticle-num}
\bibliography{QM15Refs.bib}

\begin{thebibliography}{10}
\expandafter\ifx\csname url\endcsname\relax
  \def\url#1{\texttt{#1}}\fi
\expandafter\ifx\csname urlprefix\endcsname\relax\def\urlprefix{URL }\fi
\expandafter\ifx\csname href\endcsname\relax
  \def\href#1#2{#2} \def\path#1{#1}\fi

\bibitem{Abreu20112737}
L.~M. Abreu, D.~Cabrera, F.~J. Llanes-Estrada, J.~M. Torres-Rincon, Annals of
  Physics 326~(10) (2011) 2737 -- 2772.

\bibitem{PhysRevLett.113.142301}
L.~Adamczyk, et. al., Phys. Rev. Lett. 113 (2014) 142301.

\bibitem{PhysRevLett.111.102301}
B.~Abelev, et. al., Phys. Rev. Lett. 111 (2013) 102301.

\bibitem{arXiv.1509.06888}
J.~Adam, et. al., arXiv:1509.06888.

\bibitem{v2Methods}
A.~M. Pozkanzer, S.~A. Voloshin, Phys. Rev. C 58~(1671).

\bibitem{arXiv.1212.3650}
J.~Masui, A.~Schmah, arXiv:1212.3650.

\bibitem{PhysRevLett.93.252301}
J.~Adams, et. al., Phys. Rev. Lett. 93 (2004) 252301.

\bibitem{PhysRevC.77.054901}
B.~I. Abelev, et. al., Phys. Rev. C 77 (2008) 054901.

\bibitem{PhysRevC.91.014904}
M.~Nahrgang, J.~Aichelin, S.~Bass, P.~B. Gossiaux, K.~Werner, Phys. Rev. C 91
  (2015) 014904.

\bibitem{PhysRevC.86.014903}
M.~He, R.~J. Fries, R.~Rapp, Phys. Rev. C 86 (2012) 014903.

\bibitem{PhysRevC.88.044907}
S.~Cao, G.-Y. Qin, S.~A. Bass, Phys. Rev. C 88 (2013) 044907.

\end{thebibliography}







\end{document}